\documentclass{sigcomm-alternate}
\usepackage{amsmath,amsthm,amssymb}
\usepackage{url}
\usepackage{graphicx}
\usepackage{wrapfig}

\theoremstyle{remark}

\begin{document}
\title{COMPARE: COMParative Advantage driven REsource allocation for Virtual Network Functions
}

\author{Bernardo A. Huberman, Puneet Sharma\\ \{bernardo.huberman,puneet.sharma\}@hpe.com\\
{Hewlett Packard Labs} \\
{1501 Page Mill Road}\\
{Palo Alto, CA 94304}}

\maketitle

\begin{abstract}

 As Communication Service Providers (CSPs) adopt the Network
 Function Virtualization (NFV) paradigm they need to transition
 their network function capacity to a virtualized infrastructure
 with different Network Functions running on a set of heterogeneous
 servers. This abstract describes a novel technique for allocating
 server resources (compute, storage and network) for a given set of
 Virtual Network Function (VNF) requirements. Our approach helps
 the telco providers decide the most effective way to run several
 VNFs on servers with different performance characteristics. Our
 analysis of prior VNF performance characterization on
 heterogeneous/different server resource allocations shows that the
 ability to arbitrarily create many VNFs among different servers'
 resource allocations leads to a comparative advantage among
 servers. We propose a VNF resource allocation method called COMPARE that
 maximizes the total throughput of the system by formulating this
 resource allocation problem as a comparative advantage problem among heterogeneous servers.
 There several applications for using the VNF resource allocation from COMPARE including transitioning
 current Telco deployments to NFV based solutions and providing initial VNF placement for Service Function Chain (SFC)
 provisioning.


\end{abstract}

\section{Introduction}

As Communication Service Providers (CSPs) adopt the Network Function
Virtualization (NFV)~\cite{etsi-nfv} paradigm they need to transition their network
function capacity to a virtualized infrastructure with different
Network Functions running on a set of heterogeneous servers. Given a
set of Virtual Network Function (VNF) requirements (capacity and
resource), one of the important problems being faced by the telco
providers is to decide the most effective way to run several VNFs
on servers with different performance characteristics. Efficient
resource allocation of infrastructure resources to meet VNF capacity
requirements is of utmost importance if operators are to extract the
promised NFV benefits in terms of capital and operational expenses.

\begin{figure*}
	\centering
	\includegraphics[width=0.7\textwidth]{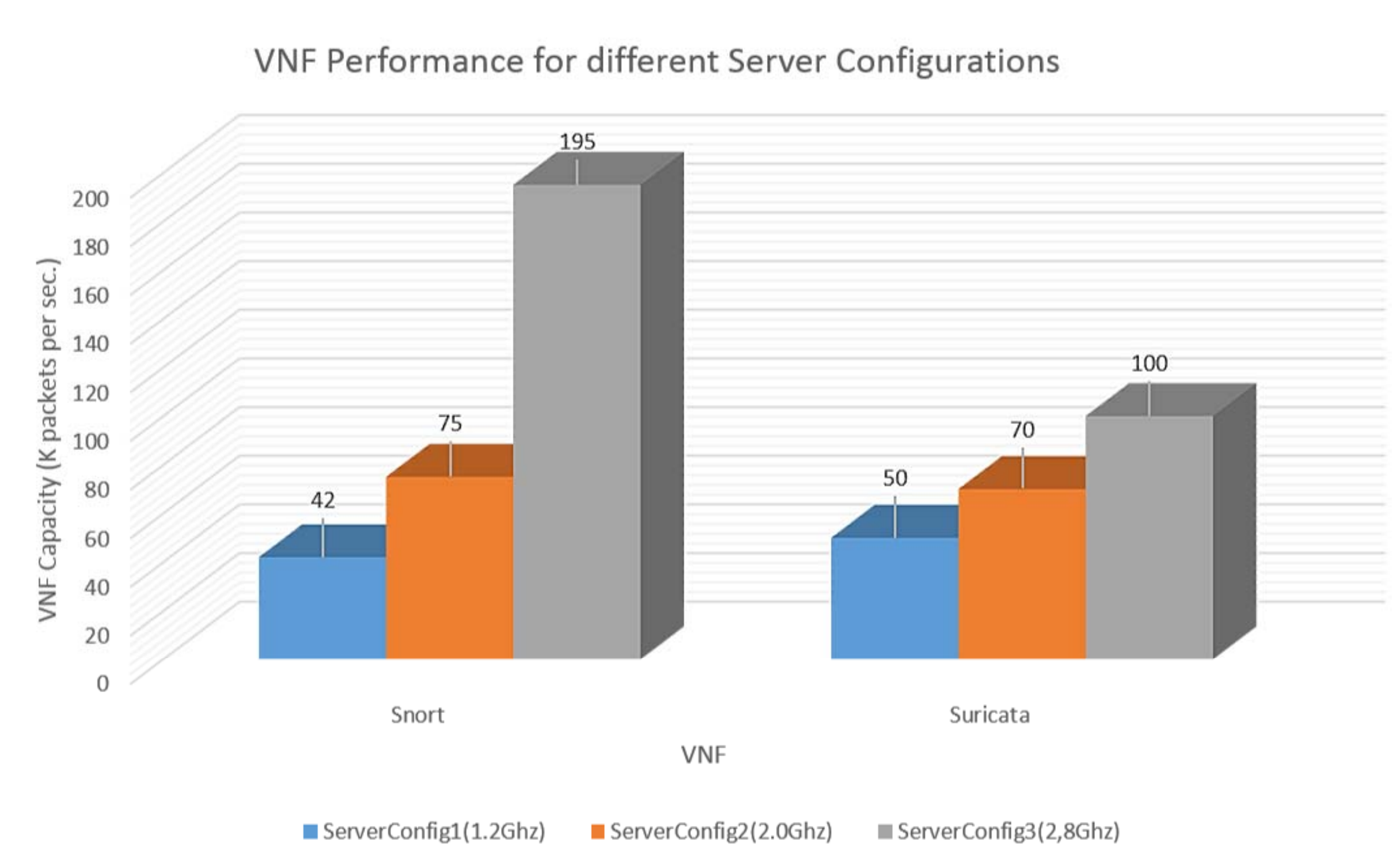}
	\caption{VNF Performance Variation on Heterogeneous Server Configs}
	\label{fig:snort-suricata}
\end{figure*}

Before discussing the problem of VNF resource allocation on a set of heterogeneous servers, we want to highlight that prior studies of VNF performance characterization on heterogeneous/different server resource allocations show that
different servers exhibit varying capacity (maximum throughput) for
different Virtual Network Functions~\cite{nfv-vital-2015}. For instance, Figure~\ref{fig:snort-suricata} shows the packet processing capacity of three heterogeneous server configurations when running two intrusion detection system (IDS) VNFs, namely, {\em Snort} and {\em Suricata}.
We used NFV-VITAL~\cite{nfv-vital-2015} tool framework to capture VNF capacity on three different server configurations. In order to emulate heterogeneous servers we artificially adjusted the CPU frequency to three different values. This was done using {\em cpufreq-set} tool that is available as part of {\em cpufrequtils} package~\cite{cpufrequtils}. Figure~\ref{fig:snort-suricata} shows that there is not only large variability in terms of VNF capacity on different server configurations but also some server configurations have absolute advantage in terms VNF performance. Our experiments indicate that such behavior is due to the difference in how different VNFs use various resources for performing the network function. This is a function of both the network function as well its implementation. In this particular case both {\em Snort} and {\em Suricata} are IDS VNFs but are implemented differently.

Such
performance and capacity variations of VNF deployment within heterogeneous
resources can be expressed in terms of a absolute advantage,
where a first resource (server) configuration has higher capacity for a
particular VNF than the second resource (server) configuration for the same
VNF. We point out that most prior resource allocation approaches leverage this absolute
advantage in allocating computational resources to different VNFs. In general, VNF resource allocation problem has been modeled as an optimization problem. For instance, the VNF orchestration problem is considered in~\cite{Bari2015}) that attempts to incorporate multiple optimization objectives such as VNF deployment costs, operating costs, penalties for service level agreement violations, and resource fragmentation costs. There are several other proposals for VNF placement and resource allocation problems that rely on similar optimization problem formulations. However, given the high computation complexity of such problems, these proposals have to invariably rely on heuristics based approaches for VNF resource allocation.

In this paper we propose a novel approach for VNF resource allocation
that exploits instead the economic principle of comparative
advantage~\cite{dfs-77}. We decribe the basics of comparative advantage in next section. As we show later in the paper, leveraging comparative advantage not only maximizes the total throughput of the system among heterogeneous servers, but also achieves near optimal allocation of server resources to different VNFs to meet the specified requirements.

Before describing our VNF resource allocation system, we discuss the basics of Comparative Advantage, which originated in the field of Economics in Section~\ref{sec:basics}. We then present the architecture of our COMPARE system  in Section~\ref{sec:system}, followed by the description of our VNF resource allocation mechanism and its optimality in Section~\ref{sec:allocation}. Section~\ref{sec:illustration} illustrates our methodology for resource allocation with a simple example of two VNFs and two server configurations. We then discuss the operation of the COMPARE system in Section~\ref{sec:operation}. Concluding remarks and future work are presented in Section~\ref{sec:conclusion}

\section{Comparative Advantage Basics}
\label{sec:basics}
In this section we discuss the basics of the comparative advantage
principle in terms of VNF performance with heterogeneous server/resource
allocations.  Consider the case of having two virtual network
functions, which we call $VNF_1$ and $VNF_2$, and two server
configurations, $machine_1$ and $machine_2$ where they can be
implemented and deployed. We say that $machine_1$ has an
\emph{absolute advantage} over $machine_2$ in one VNF if the
capacity (or maximum throughput, e.g. the number of packets processed per unit time) of $machine_1$, is higher than the capacity of running that same VNF in
$machine_2$.

A more careful capacity analysis of such a resource allocation process, shows however than in many cases $machine_2$ should only run the VNF in
which it has a \emph{comparative advantage} to $machine_1$. We say that
$machine_2$ has a comparative advantage over $machine_1$ in executing a
given VNF if the \emph{relative} throughput of $machine_2$ while running that
VNF over the other is higher than the relative throughput from running
it in $machine_1$.

This result can at times seem paradoxical, for it leads to situations
whereby although $machine_2$ can run $VNF_1$ twice more effectively
than $VNF_2$, it should only run $VNF_2$ in order to maximize the
total system throughput.

In what follows we consider the problem of resource allocation when
running \emph{multiple} VNFs on a set of heterogeneous servers taking
into account their varying processing capacity in terms of various VNFs. Given a
set of Virtual Network Function (VNF) requirements (capacity and
resource), we solve the problem of deciding the most effective way to
run several VNFs on servers with different performance
characteristics. Our system, called {\em Compare}(COMParative Advantage REsource allocation), determines the
optimal allocation of computing resources to several VNF's by
characterizing their comparative advantage.

\section{Compare: System Description}
\label{sec:system}
\begin{figure}[!th]
	\centering
	\includegraphics[width=0.5\textwidth]{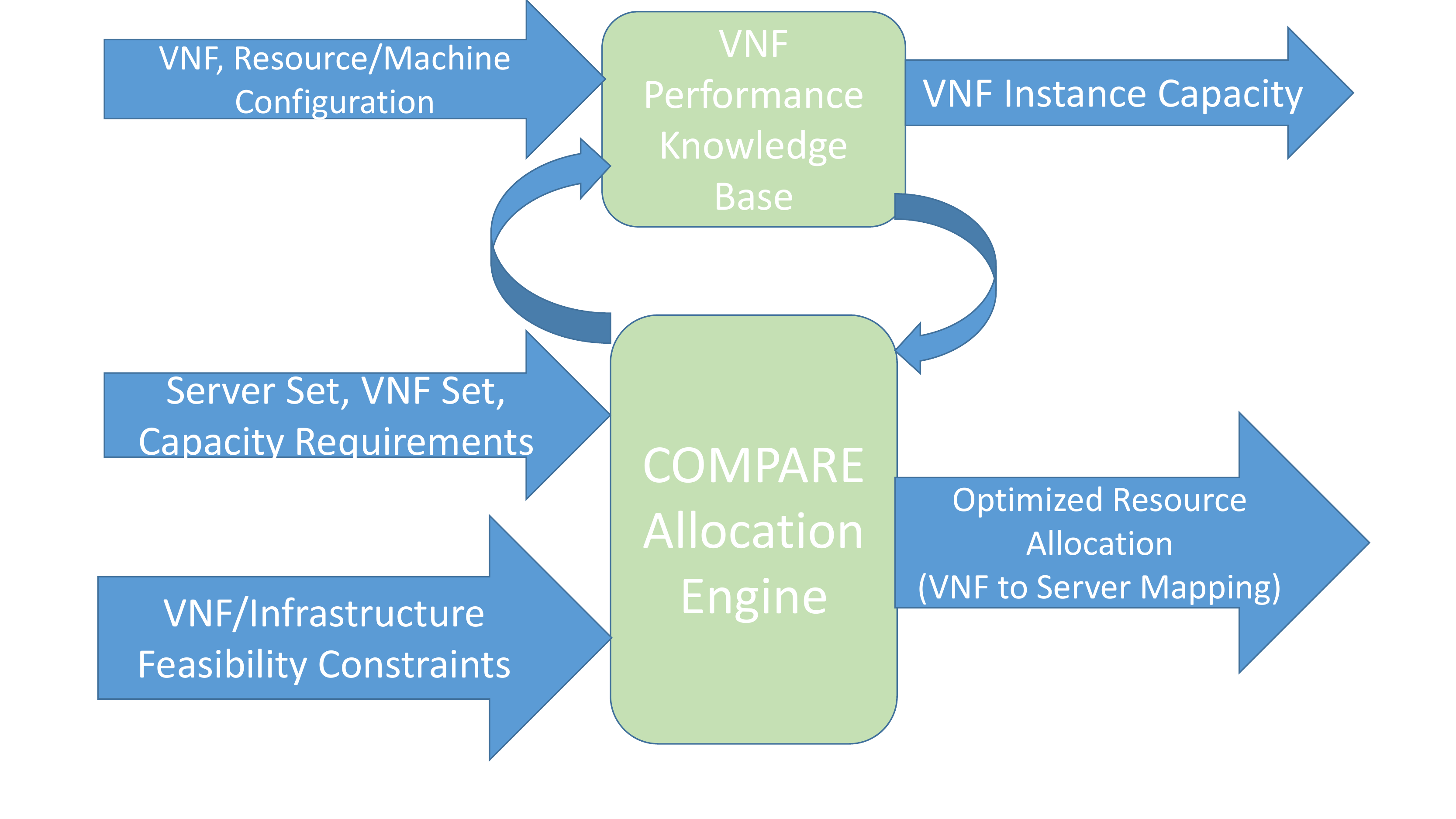}
\caption{Compare System Diagram}
\label{fig:system}
\end{figure}

Our method for deploying multiple
VNFs on a set of heterogeneous servers takes into account their varying processing capacity in terms of the VNF requirements that one wishes to deploy. Our system, called Compare(COMParative Advantage REsource allocation), determines the
optimal allocation of computing resources to several VNF's by
characterizing their comparative advantage. Figure~\ref{fig:system}
shows the COMPARE architecture, which consists of two main
components. The first component is the VNF Performance
Knowledgebase that acts as a repository of the performance characterization
data for different VNFs with different resource allocations (cpu,
memory, virtualization configuration etc.). The second component is the COMPARE
allocation engine, that solves the resource allocation optimization
problem in terms of comparative advantage and plugs-in the appropriate VNF
performance data for feasible allocation options based on available NFV Infrastructure. Besides taking the Server Set and VNF Capacity requirements as input, the operator can also provide various VNF and Infrastructure feasibility constraints such as preferred virtual slicing sizes etc.

The VNF Performance Knowledgebase is populated not only with the VNF capacity numbers for complete allocation of each available server but also the VNF capacity numbers for various permitted partial allocations of each server to the VNFs. This allows VNF Knowledgebase to capture various virtual slicing overheads. VNF characterization frameworks like NFV-VITAL~\cite{nfv-vital-2015} can be used for this purpose.

\section{Allocating resources to VNF's}
\label{sec:allocation}
\subsection{The model}

Consider a set of VNF's running in different \emph{machines}, which
may be either virtual machines, actual cores of CPU or some other
form of feasible resource allocation. Each of these machines can be
used to run the VNFs in their entirety, or virtually sliced to run several
VNFs at any given time.

Let $u$ be an $n\times m$ nonnegative matrix whose entry $u_{ij}$
represents the fraction of machine $i$ is allocated to a given VNF
$j$. For example,the machine might perform $n$ many VNFs. This
fraction can be either in absolute value or a relative one. Let $U$ be
the feasible allocation set.This feasible allocation set is
determined based on the server characteristics, available
virtualization configurations and VNF implementations. It must be noted that the feasible allocations are made available to the system based on operator's choice and considerations. If resource allocation mechanism
allocates a portion $u_{ij}$ of its $i$'th resources (e.g. processing
power) to execute one VNF $j$, the feasible allocation set is then
represented as:

\begin{equation}
  U=\left\{ (u_{ij})_{n\times m}: u_{ij}\ge 0,\ \sum_j u_{ij}\le 1
  \right\}.
\end{equation}

Let $x$ be a non-negative $m$-vector whose $i$'th component $x_i$
measures the number of packets that are processed in a given
time. Note that while packets processed per unit time is one of the
metrics to measure VNF performance, our approach can work with other
performance/throughput metrics as well.

\begin{equation}
  \label{eq:x-u}
  x_j = b_{1j}u_{1j} + \cdots + b_{nj}u_{nj}, \quad j=1, \dots, m,
\end{equation}.

Thus $b_{ij}$ measures the effectiveness of machine $i$ at running VNF
$j$, which is again measured in terms of the number of packets processed per
unit time. It is clear from Equation~\ref{eq:x-u} that in this section, we assume the overhead associated with virtual slicing of server resources to be zero. While this allows us to provide a clean proof for optimality of comparative advantage based allocation, we later demonstrate that non-zero virtualization overhead does not invalidate the optimality of our solution.

What our approach does is to attempt to optimize the system's utility. The
utility function can be defined in different ways depending on
operators preferences. For instance, the operator's utility function
can be expressed as the gain obtained by the revenue generated by running a
given set of VNFs minus its infrastructure costs:

\begin{equation}
V=g(x)-c(u).
\end{equation}

In situations where the cost is a constant one can write
$V=g(x)$, where $g$ is a pay-off function which is strictly increasing
in $x$.


 We also make a technical assumption that $g$ satisfies the Inada
 conditions:
\begin{equation}
\lim_{x_j\to 0} \frac{\partial g(x)}{\partial x_j} =\infty \quad
\text{for all }j=1,\dots, m.
\end{equation}

Thus the our approach seeks to solve the following optimization problem:

\begin{equation}
  \max\, g(x) \quad \text{s.t. } u\in U.
\end{equation}

\subsection{Leveraging comparative advantage}

In this section we describe how to derive the resource allocation for a given set
of VNFs and infrastructure resources. We begin using comparative advantage for the simple case of allocating two VNFs to two machines.

\subsubsection{\label{sec:2m2j} Two machines and two VNFs}

Let us start with the simplest case: there are only two machines available
and two VNFs ($m=n=2$). The objective is to maximize the overall system utility which is a function of
the total number of packets processed in a given time as described earlier, or equivalently

\begin{equation}
  \label{eq:problem-2-2}
  \begin{array}{ll}
    \max & g(x_1, x_2) =
g(b_{11}u_{11}+b_{21}u_{21}, b_{12}u_{12}+b_{22}u_{22})\\ \text{s.t. }
& u\ge 0, \ u_{11} + u_{12} \le 1, \ u_{21} + u_{22} \le 1.
 \end{array}
\end{equation}

We say that $machine_1$ has \emph{comparative advantage} for
running $VNF_1$ if
\begin{equation}
  \label{eq:comparative-advantage}
  \frac{b_{11}}{b_{21}} > \frac{b_{12}}{b_{22}}.
\end{equation}

Clearly, under this definition $machine_2$ has a comparative advantage
over $machine_1$ for running $VNF_2$. This result can seem to be
counter-intutive in some cases. For example, consider the case where
$b_{11}=5$, $b_{12}=b_{21}=2$, and $b_{22}=1$. Although $machine_2$
can perform $VNF_1$ two times more efficiently than $VNF_2$, it should
only perform $VNF_2$.

From Eq.~(\ref{eq:comparative-advantage}) we can show that
either $u_{12}=0$ or $u_{21}=0$ in the optimal allocation. Suppose
otherwise that both $u_{12}>0$ and $u_{21}>0$. Consider the
following small change in $u$:

\begin{equation}
  \Delta u_{11} = -\frac{b_{21}}{b_{11}} \Delta u_{21} = - \Delta u_{12} = \frac{b_{22}}{b_{12}} \Delta u_{22} >0.
\end{equation}

When the change is small we can keep $u_{12}>0$ and $u_{21}>0$.  The
value of $g$ will not be affected since $x_1$ and $x_2$ remain
unchanged. It is easy to check that while the first constraint in
Eq.~(\ref{eq:problem-2-2}) is binding after the change, the second
constraint cannot be satisfied, i.e.:

\begin{equation}
  \Delta u_{21} + \Delta u_{22} = -\frac{b_{22}}{b_{12}} \left(\frac{b_{11}}{b_{21}} - \frac{b_{12}}{b_{22}} \right) \Delta u_{12} <0.
\end{equation}

Thus one can increase both $x_1$ and $x_2$ without violating the
constraints, but doing so will cause an increase in $g$ and
contradict optimality. Therefore, it cannot be that both
$u_{12}>0$ and $u_{21}>0$; one of them has to be zero.

When $u_{12}=0$ $machine_1$ performs only $VNF_1$, so $u_{11}>0$.
It follows from the Inada condition that $VNF_2$ has to be
performed by $machine_2$, because the profit margin at $x_2=0$ is
infinity. Thus $u_{22}>0$. When $u_{21}=0$ a similar argument leads
to the same conclusion, i.e.~$u_{11}>0$ and $u_{22}>0$. This means
that if a machine has comparative advantage in performing a given VNF
then it should always run that VNF (it may or may not
run the other VNF). This depends on the capacity requirements of the operator for each VNF.

\bigskip

We list without proof the optimal solution for three possible
cases, neglecting degeneracy:

\medskip \noindent \emph{Case 1.} $\frac{b_{11}}{b_{21}} > \frac{b_{12}}{b_{22}}> 1$.
\begin{equation}
u_{11}=\frac{b_{11}}{b_{12}} \frac{b_{12}+b_{22}}2,\quad
u_{12}=\frac{b_{12}-b_{22}}2,\quad u_{21}=0,\quad u_{22}=b_{22}.
\end{equation}
\medskip

Again, our result says that if $machine_1$ has \emph{absolute advantage} over
$machine_2$ in both VNF functions, then $machine_2$ should only
perform the function in which it has comparative advantage. This
result can be counter intuitive and perplexing in some cases. For
example, consider the case where $b_{11}=5$, $b_{12}=b_{21}=2$, and
$b_{22}=1$. Although $machine_2$ can execute $VNF_1$ two times more
effectively than $VNF_2$, it should only execute $VNF_2$.

\medskip \noindent \emph{Case 2.} $1 > \frac{b_{11}}{b_{21}} > \frac{b_{12}}{b_{22}}$.

\begin{equation}
  u_{11}=b_{11},\quad u_{12}=0,\quad u_{21}=\frac{b_{21}-b_{11}}2,
\quad u_{22}=\frac{b_{22}}{b_{21}} \frac{b_{11}+b_{21}}2.
\end{equation}

It can be noted from above equation that this is similar to Case 1.

\medskip \noindent \emph{Case 3.} $\frac{b_{11}}{b_{21}} > 1 > \frac{b_{12}}{b_{22}}$.

\begin{equation}
  u_{11}=b_{11},\quad u_{12}=u_{21}=0,\quad u_{22}=b_{22}.
\end{equation}

In other words, both $machines_{1,2}$ should specialize if and only
if each machine has absolute advantage in executing one particular VNF.

\subsubsection{The comparative advantage generalization}

The result of Section \ref{sec:2m2j} can be generalized to the
case of more than two machines and more than two VNF functions. Assume
that
\begin{equation}
  \frac{b_{i_1j_1}}{b_{i_2j_1}} >\frac{b_{i_1j_2}}{b_{i_2j_2}}
\end{equation}

for machines $i_1, i_2$ and VNF's $j_1, j_2$. In this cae, it follows that one of
$u_{i_1j_2}$ and $u_{i_2j_1}$ must be zero.

\subsubsection{Two machines and $m$ VNFs}

Using the above generalization let us now consider the case of allocating $m$ VNFs on two machines. Without loss of generality we can order the machines by
comparative advantage, so that $machine_1$ has comparative advantage
in performing functions with smaller labels:

\begin{equation}
  \frac{b_{11}}{b_{21}} > \cdots > \frac{b_{1m}}{b_{2m}}.
\end{equation}

By the comparative advantage generalization, for any $1\le j < k
\le m$ it must be that either $u_{2j}=0$ or $u_{1k}=0$. Therefore
there must exist some $J$ such that

\begin{equation}
  \label{eq:ca-characterization}
  \begin{array}
    {ll} u_{1j}>0, \quad
    u_{2j}=0 & \text{for } 1\le j <J; \\
    u_{1j}=0, \quad u_{2j}>0 & \text{for } J < j \le m.
  \end{array}
\end{equation}

In words, $machine_1$ should execute functions (VNFs) $1, \dots, J-1$ and
possibly $J$, and $machine_2$ should perform functions $J+1, \dots,
m$ and possibly $J$. Once again, it must be noted that allocation thresholds are determined by operator's specification of required capacity for each VNF.

\subsubsection{$n$ machines and two VNF functions}

Once again we label the machines in decreasing order of their
comparative advantage:
\begin{equation}
  \frac{b_{11}}{b_{12}} > \cdots > \frac{b_{n1}}{b_{n2}}.
\end{equation}

Like before, the solution has a simple form
\begin{equation}
  \begin{array}
    {ll} u_{i1}>0, \quad u_{i2}=0 & \text{for } 1\le i<I; \\
    u_{i1}=0, \quad u_{i2}>0 & \text{for } I < i \le n.
  \end{array}
\end{equation}

In words, machines $1, \dots, I-1$ should execute $VNF_1$ and
machines $I+1, \dots, n$ should execute $VNF_2$. Machine $I$
may perform both functions.

\bigskip

From the comparative advantage generalization, the optimal machine allocations have the simple form
  \begin{eqnarray}
    x_1 &=& b_{11} + \cdots b_{I-1,1} + \theta b_{I1},\\
    x_2 &=& (1-\theta) b_{I2} + b_{I+1, 2} + \cdots + b_{n2},
  \end{eqnarray}
  where $0\le \theta < 1$. Ignoring degeneracy for
the moment (i.e.~assuming that $0<\theta<1$), the optimal $\theta$
must satisfy the first order condition
\begin{equation}
  \label{eq:foc}
  \frac{\partial \log g(x)}{\partial \theta} = \frac \alpha {x_1} b_{I1} - \frac {1-\alpha}{x_2} b_{I2} = 0,
\end{equation} or

\begin{equation}
  \label{eq:cd-solution}
  \frac{\alpha \, b_{I1}}{(1-\alpha)b_{I2}} = \frac{b_{11} + \cdots + b_{I-1,1} + \theta b_{I1}}{(1-\theta) b_{I2} + b_{I+1, 2} + \cdots + b_{n2}}.
\end{equation}

This equation has a solution since the left side decreases with
$I$ and the right side increases with $I$.

\begin{figure*}
	\centering
	\includegraphics[width=0.9\textwidth]{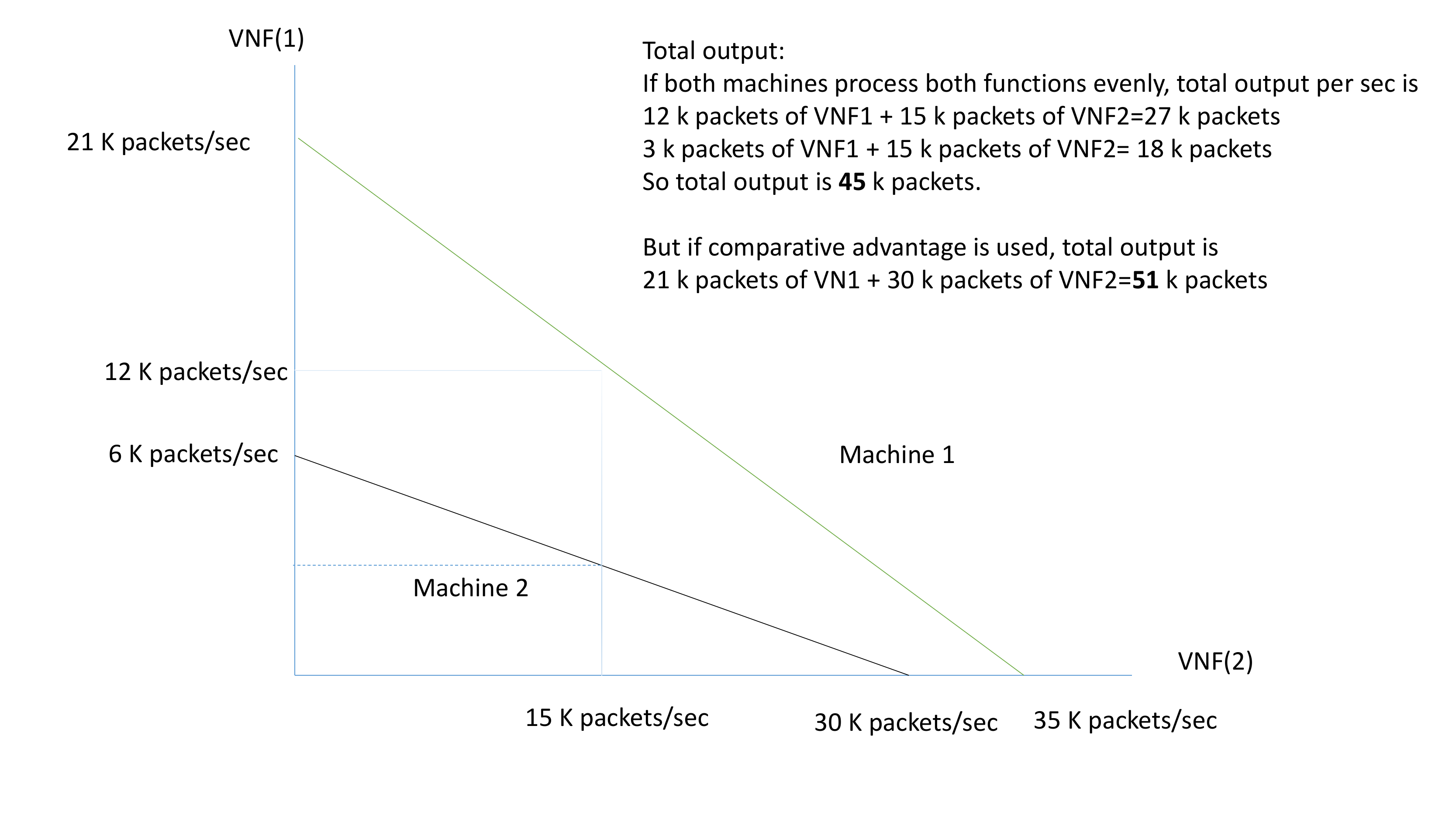}
	\caption{Resource allocation for two VNF's among two machines}
	\label{fig:twomac-twovnf}
\end{figure*}

If we define two shadow prices
\begin{equation}
\label{eq:cd-price-ratio} p_1=\frac{\partial \log g(x)}{\partial
x_1} = \frac \alpha {x_1}, \quad p_2=\frac{\partial \log
g(x)}{\partial x_2} = \frac {1-\alpha} {x_2}, \end{equation}
Eq.~(\ref{eq:foc}) can be also written as $p_1 b_{I1}=p_2 b_{I2}$,
so machine $I$ is indifferent to performing $VNF_1$ or $VNF_2$.

Note that in the general case one can no longer sort the machines
or functions by comparative advantage, and has to solve the full
optimization problem. The comparative advantage characterization
still holds though.

\section{COMPARE Illustration}
\label{sec:illustration}
In the earlier section we presented analytical proof of the optimality of COMPARE's comparative advantage based resource allocation approach.

Figure~\ref{fig:twomac-twovnf} illustrates the COMPARE resource allocation mechanism with a simple example of two machines and two VNFs. Machine 1 can process 21Kpps of VNF(1) and 35Kpps of VNF(2). Similarly, Machine 2 can process only 6Kpps of VNF(1) and 30Kpps of VNF(2). Assuming zero virtual slicing overhead the packet processing capacity for partial allocation of machine resources to a VNF can be represented by joining the two VNF numbers for 100\% allocation of the machine. It can be seen from the figure that evenly distributing both the machine resources to the two VNFs can only process total of 45Kpps while comparative advantage based allocation can process total of 51Kpps. It is evident, the comparative advantage based resource allocation achieves higher overall system throughput than one based on absolute advantage based allocation.

Based on our prior work of VNF characterization~\cite{nfv-vital-2015}
we have started building VNF characterization Knowledgebase for various VNFs as shown in Figure~\ref{fig:system}. We are collecting characterization information for different IDS VNFs such as Snort~\cite{snort}, and Suricata~\cite{suricata} as well as IMS VNF like Clearwater~\cite{clearwater} etc. Preliminary performance characterization of IDS VNFs {\em Snort} and {\em Suricata} on two different machine configurations shown in Figure~\ref{fig:snort-suricata} exhibit results similar to those of Figure~\ref{fig:twomac-twovnf}.

\begin{figure}
	\centering
	\includegraphics[width=0.5\textwidth]{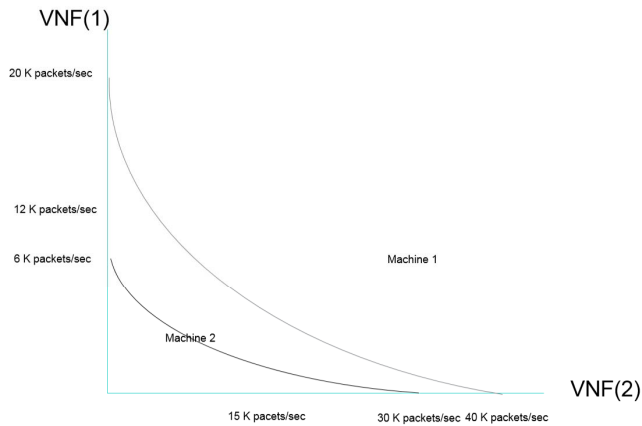}
	\caption{Performance curve for VNFs to illustrate Virtual Slicing Overhead of Machines}
	\label{fig:virtoverhead}
\end{figure}

We now look what the impact of considering virtual slicing overhead on our VNF resource allocation mechanism. Figure~\ref{fig:virtoverhead} illustrates the performance characterization curve for the two VNFs when the virtual slicing overhead is non-zero. It is evident that, in this case, while virtual slicing overhead can dversely impact the overall system throughput for absolute advantage based allocation, system throughput for comparative advantage allocation still performs better. As we mentioned earlier, such virtual slicing overheads are capture in the VNF performance knowledgebase component of COMPARE architecture.

\section{COMPARE operation}
\label{sec:operation}
The COMPARE system relies on VNF characterization information that captures the effectiveness of available servers (and feasible configurations) for implementing different VNFs of interest to the operator. VNF performance characterization frameworks like NFV-VITAL~\cite{nfv-vital-2015} can be used to populate the VNF Characterization Knowledgebase shown in our system architecture diagram. It must be noted that system issues such as performance degradation due to virtualization or performance impact of resource sharing are captured appropriately by the VNF Knowledgebase. For a given set of VNFs, available server resources and feasible configurations, the COMPARE resource allocation engine creates a comparative advantage based model by querying VNF Knowledgebase for different $b_{ij}$ values for various feasible configurations.

Once the model is parameterized,the COMPARE resource allocation engine can sort machines such that lower-indexed machines have comparative advantage in performing VNFs with smaller labels and thus provide optimal resource allocation. In the general case, COMPARE can be implemented using various comparative advantage planners such as the one described in~\cite{little-69}.

Though the focus of this paper is on using COMPARE architecture for VNF resource allocation for optimizing overall system throughput, there are several other applications of our mechanism. For instance, comparative advantage based allocation can be used as heuristic for faster VNF placement approaches for Service Function Chaining (SFC) that consider end-to-end SFC latency~\cite{Chua2016}. Similar COMPARE approach can be leveraged by operators for performing cost-benefit analysis of migrating their current network function deployments to NFV based infrastructure.

\section{Conclusion}
\label{sec:conclusion}

In this paper we have shown how an approach based on comparative advantage can lead to optimal allocation of VNF's among a set of heterogenous servers. We did so by describing the basic idea of comparative advantage, a well established principle in trade economics, and showed how it leads to optimal allocation. Furthermore we described the COMPARE architecture which does the actual deployment of our approach.  While existing approaches have taken the seemingly obvious course of using absolute advantage to decide on how to deploy VNFs, we showed how comparative advantage is a much better solution, as it computes the opportunity cost of deployments in different platforms.Given the present trend towards virtualization of most network functions, and the fact that they are deployment among servvers with different characteristics, it is crucial to decide on optimal allocatins. The COMPARE approach offers such a solution.

\section*{Acknowledgements}
The authors would like to express their thanks to our Hewlett Packard Labs intern, Lianjie Cao, Ph.D. Student, Purdue University for carrying out the IDS VNF performance characterization using NFV-VITAL framework.

\end{document}